\documentclass[prd,twocolumn,notitlepage,amsmath,nofootinbib,superscriptaddress,showpacs,showkeys]{revtex4-1}

\usepackage[brazil,english]{babel}
\usepackage{amsmath}
\usepackage{graphicx}
\usepackage{amssymb}
\usepackage{esint}
\usepackage[usenames,dvipsnames]{color}
\usepackage{longtable}
\usepackage{dcolumn}
\usepackage{ulem}
\usepackage{nicefrac}

\begin{document}

\title{Self-similar resistive circuits as fractal-like structures}

\author{Claudio Xavier Mendes dos Santos}
\email{claudioxm@gmail.com}
\affiliation{Instituto de Ci\^{e}ncias Matem\'{a}ticas e de Computa\c{c}\~{a}o, Universidade de S\~{a}o Paulo,\\ Av. Trabalhador S\~{a}o Carlense 400, 13566-590, S\~{a}o Carlos, Brazil}

\author{Carlos Molina Mendes}
\email{cmolina@usp.br}
\author{Marcelo Ventura Freire}
\email{mvf@usp.br}
\affiliation{Escola de Artes, Ci\^{e}ncias e Humanidades, Universidade de S\~{a}o Paulo,\\ Av. Arlindo Bettio 1000, 03828-000, S\~{a}o Paulo, Brazil}

\begin{abstract}

Fractals play a central role in several areas of modern physics and mathematics. 
In the present work we explore resistive circuits where the individual resistors are arranged in fractal-like patterns. These circuits have some of the characteristics typically found in geometric fractals, namely self-similarity and scale invariance.
Considering resistive circuits as graphs, we propose a definition of self-similar circuits which mimics a self-similar fractal. 
General properties of the resistive circuits generated by this approach are investigated, and interesting examples are commented in detail. Specifically, we consider self-similar resistive series, tree-like resistive networks and Sierpinski's configurations with resistors.
\end{abstract}

\keywords{self-similarity, resistive circuit, fractal}

\maketitle

\section{\label{intro} Introduction} 

Fractality is a relatively recent but powerful concept that permeates much of physics and mathematics. 
Fractals can be a powerful teaching tool \cite{Camp, Portes,Amaku,Elvia,Assis,Caicedo}, and formal and rigorous investigation of fractal systems are cutting edge mathematics \cite{mandelbrot2004fractal}. At the same time, fractals are used in realist modelings of nature. Quoting Beno\^{\i}t Mandelbrot \cite{mandelbrot1983fractal}, %
``Clouds are not spheres, mountains are not cones, coastlines are not circles, and bark is not smooth, nor does lightning travel in a straight line.''

There are several, but not quite equivalent, precise definitions for the modern idea of a fractal \cite{mandelbrot1983fractal,Assis}. Informally, a basic characteristic of this structure is scale invariance, that is, the fractal pattern remains the same with any magnification. This property is denoted as self-similarity. 
A common way to construct self-similar structures is to define them recursively, using a set of iterated functions \cite{kirillov2013tale}. Several popular self-similar geometric figures can be constructed in this way, such as the fractal tree, the Sierpinski triangle and the Cantor set \cite{kirillov2013tale}.

Networks and circuits that have fractal-like patterns typically have an infinite number of elements. Infinite circuits have extensive physical and mathematical applications \cite{zemanian,atkinson,cserti,hansen:1986,chen2016,ruiz2017}.
For instance, physics oriented works involving self-similar resistive circuits include the investigation of resistance properties of Sierpinski triangle fractal networks \cite{bedrosian,boyle,Alstrom198820} and binary tree circuits \cite{Essoh:1988bn}, among several other fractal patterns. 
More rigorous mathematical aspects of the issue were also developed \cite{boyle,Essoh:1988bn}.

In the present work, a general definition for self-similar resistive circuits is proposed and explored. Analyzed from the point of view of the Graph Theory, resistor networks are recursively constructed. In our approach, the issue of calculating the equivalence resistance of a self-similar circuit is transformed into a fixed point problem. Interesting particular cases are treated in detail. 

The structure of this paper is presented in the following. 
In section~\ref{graphs} we introduce the notation used in the paper and review the basic results in the theory of resistive circuits, presenting those objects as graphs. 
In section~\ref{general-definition}, a general definition for self-similar circuits is proposed and explored. We derive a condition to characterize the calculation of the equivalent resistance of a self-similar circuit as a fixed-point problem.
In section~\ref{series} we consider one of the simplest self-similar circuits, the self-similar resistive series, a recursive configuration formed by resistors in series.
In section~\ref{tree-circuits} we explore self-similar resistive trees, resistive circuits analogous to fractal trees.
In section~\ref{Sierpinski-circuits} self-similar Sierpinski resistive circuits, generated with resistors arranged in self-similar Sierpinski-like configurations, are discussed. 
Final comments are presented in section~\ref{final-comments}.

\section{\label{graphs} Resistive circuits as graphs}

The building blocks of the structures presented in this work are ideal resistors, linked to each other by ideal connectors. Ideal resistive circuits constructed in this way are a simplification of real electric systems. Still, ideal circuits can be a very good approximation of real circuits in many scenarios of interest \cite{nilsson2008electric}.

An ideal resistor is a two-terminal element characterized by an electric resistance. In the analysis of ideal resistive circuits, the only relevant informations are the individual resistances of each resistor and the pattern of interconnection among them. Other aspects that are important for more realistic characterization of a physical circuits, such as the length and shape of the connectors and the positions of the resistors, are ignored in the idealized description. Another way to express this latter remark is to say that only topological information are used in the definition of an ideal resistive circuit. Geometric aspects are not relevant in the construction of these models. That feature indicates that ideal circuits are naturally represented as graphs.

A simple undirected graph $G$ is composed of a set $V=V(G)$ of vertices and a set $E=E(G)$ of edges. In such graphs, edges are ``links'' between two vertices. More precisely, the set $E$ of edges are constructed as $2$-element subsets of the set $V$ of vertices. 
Hence, the graph $G$ is the ordered pair $G=(V,E)$. We will consider that the graphs used in this work are finite, with $V$ and $E$ finite sets \cite{harary1969graph}.
Also, the graphs of interest here will be connected. That is, for each each pair of vertices in a connected graph, one can find a path from one vertex to the other \cite{harary1969graph}.
Another layer of structure that is needed are weights. A weighted graph is a graph $G=(V,E,W)$ in with a number is associated to each edge by a weight function $W$ \cite{harary1969graph}. 

Using the language of Graph Theory, a resistive circuit has a natural description in terms of graphs. A circuit is equivalent to an element of $\mathcal{C}$, where $\mathcal{C}$ is the set of simple graphs that are finite and connected, with the number of vertices equal to or greater than $2$, having strictly positive real numbers as weights.
We identify the individual resistors in the circuit with the graph edges, and the connection points of the resistors with the graph vertices. The individual resistances of each resistor are codified in the weights. In the present work, we consider ideal resistive circuits as graphs. Fixing terminology, along this manuscript both terms (``circuits'' and ``graphs'') will be considered as associated to the same objects. 

\begin{figure}[h]
\centering

\includegraphics[width=\columnwidth]{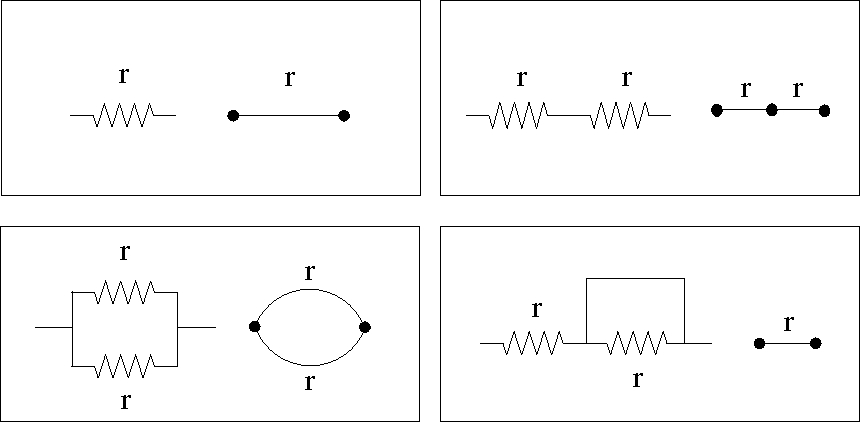}

\caption{Diagrammatic representations for several resistive circuits, using a more traditional notation (left in each box), and the more compact graph notation (right in each box).}
\label{graphic-notation}
\end{figure}

As an additional benefit, graph language gives us a compact way to represent and manipulate resistive circuits.
Diagrammatically we will represent a circuit, understood as a graph, by dots and links. When convenient, individual resistances (weighs) will be included in the diagram. 
This notation is similar to the more usual notation for resistive circuits \cite{nilsson2008electric}. 
As an illustration, in figure~\ref{graphic-notation} we compare the common graphical representations for resistive circuit with the diagrammatics suggested here.
But it is not always a matter of substituting conventional resistor symbols by plain lines. For instance, in the usual notation, short circuit connections are allowed. In plain graph notation, sections of a resistor network that are short circuited have the same electric potential, and therefore are collapsed into a single vertex (see bottom right diagrams of figure~\ref{graphic-notation}).

An important issue concerning resistive circuits is the calculation of their equivalent resistance. Given a circuit with two highlighted vertices (the external terminals), Th\'evenin's and Norton's theorems imply that this circuit is electrically equivalent to a single resistor \cite{nilsson2008electric}. 
In other words, there exists a function $R_{AB}$, 
\begin{gather}
R_{AB}: \mathcal{C} \times V(\mathcal{C}) \times V(\mathcal{C})  
\rightarrow \mathbb{R}^{+}_{*} \nonumber \\
R_{AB} = R_{AB} (C) \, ,
\label{def_R}
\end{gather}
that associates a circuit $C$ and two different vertices (elements $A$ and $B$ of $V(C)$ with $A \ne B$) to elements of $\mathbb{R}^{+}_{*}$, where $\mathbb{R}^{+}_{*} = (0,\infty)$ denotes the set of strictly positive real numbers.
The number $R_{AB}(C)$ is the equivalent resistance of the circuit $C$, considering $A$ and $B$ as external vertices.

The concept of equivalent resistance implies an equivalence relation on the set of circuits. To fix notation, this relation (of having the same equivalent resistance) is indicated by $\sim$~. Considering $C_{1},C_{2} \in \mathcal{C}$, $A,B \in V(C_{1})$ and $D,F \in V(C_{2})$, we have that 
\begin{equation}
(C_{1},A,B) \sim (C_{2},D,F) \Longleftrightarrow R_{AB}(C_{1}) = R_{DF}(C_{2}) \,\,.
\end{equation}

\begin{figure}[h]

\centering
\includegraphics[width=0.6\columnwidth]{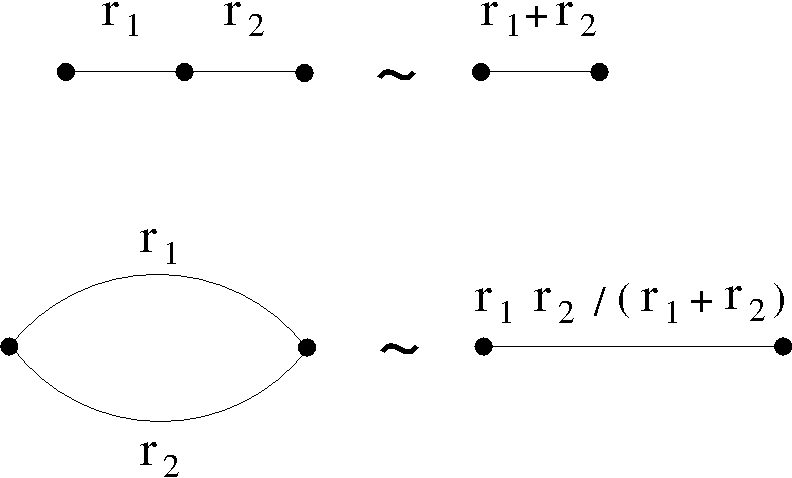}

\caption{Diagrammatic representations for the series and parallel transformations (left and right diagrams, respectively).}

\label{series-parallel}
\end{figure}

\begin{figure}[h]
\centering

\includegraphics[width=0.6\columnwidth]{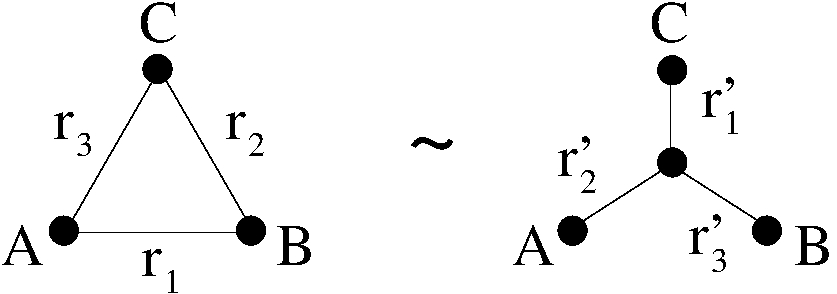}

\caption{Diagrammatic representation for the $Y-\Delta$ transformation. In the correspondence, $r'_{1} = r_{2} r_{3} / (r_{1} + r_{2} + r_{3})$, $r'_{2} = r_{1} r_{3} / (r_{1} + r_{2} + r_{3})$ and $r'_{3} = r_{1} r_{2} / (r_{1} + r_{2} + r_{3})$.}
\label{Ydelta}
\end{figure}

There are several techniques used in the calculation of equivalent resistances. A practical method involves the use of ``transformations'' in the graphs, substituting a given subgraph for an equivalent and simpler form. 
For instance, two elementary transformations are the series and parallel resistor associations \cite{nilsson2008electric}.
Using the graph notation, we illustrate those transformations in figure~\ref{series-parallel}. 
A more elaborate substitution is the so-called $Y-\Delta$ transform \cite{nilsson2008electric}, shown in figure~\ref{Ydelta}.
We will extensively apply series, parallel and $Y-\Delta$ transformations in sections~\ref{series}, \ref{tree-circuits} and \ref{Sierpinski-circuits}.

\section{\label{general-definition} Self-similar resistive circuits in general}

To consistently implement the idea of a ``fractal-like resistive circuit'', it is not enough to just take a geometric fractal and substitute lines by resistors. In a circuit, only topological characteristics are relevant. The strategy to properly define a circuit with fractal properties must be more elaborate.
In the present paper, we will consider self-similarity as our guiding concept. A geometric fractal can be defined as the limit of a recursive sequence, in such a way that its pattern is self-similar \cite{kirillov2013tale}.  
That construction will be adapted here to circuits formed by resistors, and therefore the objects defined will be denominated self-similar resistive circuits.

Following the usual construction for self-similar fractals, the key element in our definition of self-similarity is a recurrence function  $F_{AB}$, which maps a circuit with two external terminals $A$ and $B$ into another circuit with the same external terminals,
\begin{gather}
F_{AB}: \mathcal{C} \rightarrow \mathcal{C} \nonumber \\
F_{AB} = F_{AB} (C)
\,\, ,
\end{gather}
with $A, B \in V(C)$ and $A \ne B$.

Given $A$, $B$, $F_{AB}$ and a specified resistive circuit $C_{0}$ (an ``initial circuit''), a sequence of circuits $(C_{i})$, 
\begin{equation}
(C_{i}) = \left( C_{0}, C_{1}, C_{2}, \ldots \right) \,\, ,
\end{equation}
can be constructed with $F_{AB}$ in a recursive way as
\begin{equation}
C_{i+1} = F_{AB} (C_{i}) \,.
\label{recorrenceC}
\end{equation}
It should be noticed that $F_{AB}$ is not only a function of a circuit, but also of two vertices ($A$ and $B$ for example), interpreted as external terminals of the given circuit. This is necessary since the equivalent resistance of a circuit depends on the choice of external vertices.

In fact, with the function $R_{AB}$ introduced in equation~(\ref{def_R}), we associate a strictly positive number $R_{i}$ to each element of the sequence $(C_{i})$,
\begin{equation}
R_{i} = R_{AB} (C_{i}) \, ,
\end{equation}
with $R_{i} \in \mathbb{R}^{+}_{*}$.
The number $R_{i}$ is the equivalent resistance of the circuit $C_{i}$ with external terminals $A$ and $B$.
Related to the sequence of graphs $(C_{i})$, we have a sequence of real numbers $(R_{i})$,
\begin{equation}
(R_{i}) = \left( R_{0}, R_{1}, R_{2}, \ldots \right) \, .
\end{equation}

In an ideal resistive circuit, the most simple and important physical observable is the equivalent resistance. Therefore, a necessary condition for the objects treated here to be physically reasonable is that their equivalent resistance should be well-characterized.
Hence, we define as a self-similar circuit the sequence $(C_{i})$ of resistive circuits whose associated sequence $(R_{i})$ of resistances converges to a finite and non-zero equivalent resistance $R_{eq}$,%
\footnote{It should be mentioned that this condition is not universally adopted in the definition of a self-similar resistive circuit.}  
\begin{equation}
R_{eq} = \lim_{i \rightarrow \infty} R_{AB} (C_{i}) \, .
\end{equation}
As a more technical remark, we specify the usual metric $d(x,y) = |x-y|$ to characterize convergence in the sequence $( R_{i} )$ of positive real numbers. 
It should be stressed that a self-similar circuit is identified with the whole sequence of graphs (circuits) whose elements are recursively constructed. The essential information that defines a particular circuit is then an initial circuit and a recurrence function.

One way to ensure convergence of the sequence $(R_{i})$ is to find a recursive function $T$ for this sequence,
\begin{equation}
R_{i+1} = T (R_{i}) \,.
\label{recorrence-T}
\end{equation}
Following this strategy, the issue is transformed in a fixed point problem \cite{haaser1991real}. If the function $T$ has a fixed point, the sequence $(R_{i})$ converges.
Since the set $\mathbb{R}^{+}_{*}$ equipped with the usual metric $d(x,y)$ forms a complete metric space \cite{haaser1991real}, Banach fixed-point theorem \cite{haaser1991real} guarantees the existence and uniqueness of a fixed point $R_{eq}$ for $T$ if there exists a constant $\lambda$ such that 
\begin{equation}
d(T(x),T(y)) \leq \lambda d(x,y) 
\,\,\, \textrm{with} \,\,\, 0 \le \lambda < 1 \, ,
\label{contraction}
\end{equation}
for all $x$, $y$ in $\mathbb{R}^{+}_{*}$. A map which satisfies condition~(\ref{contraction}) is called a contraction \cite{haaser1991real}.
Moreover, if $T$ is a contraction, the equivalent resistance $R_{eq}$ is the solution of the fixed-point equation
\begin{equation}
R_{eq} = T(R_{eq}) \, .
\label{Req}
\end{equation}

So we have a method to check if the sequence $(C_{i})$ of resistive circuits is a self-similar circuit, in the sense proposed here. The answer is affirmative if we can find a contraction $T$ such that equation~(\ref{recorrence-T}) is satisfied. In this case, the equivalent resistance of the self-similar resistive circuit is well defined, given by the (unique) solution of equation~(\ref{Req}).

The general characterization of self-similar resistive circuits presented here will be explored in specific examples in the next sections.
We will introduce and study self-similar resistive series, self-similar resistive trees and Sierpinski resistive circuits.

\section{\label{series} Self-similar resistive series}

One of the simplest candidate for a self-similar circuit is a collection of resistors arranged in series. But in a self-similar circuit, the number of resistors grows as each new circuit is generated for the sequence. So, for the equivalence resistance to be finite in a collection of resistors arranged in series, it is not enough to consider an array of resistors with the same resistance. The construction must be more elaborate.

\begin{figure}[h]

\centering
\includegraphics[width=0.4\columnwidth]{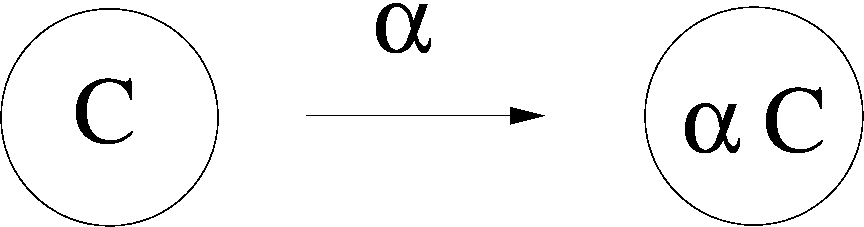} 

\caption{Diagrammatic representation for the multiplication of a circuit $C$ by the constant $\alpha$.}
\label{multiplication}
\end{figure}

Let us define the multiplication of a weighted graph $C$ by a constant $\alpha$. Informally, a graph $\alpha C$ is a graph obtained from $C$, with the same vertices and links, but with the weights multiplied by $\alpha$. 
More precisely, with definitions presented in section~\ref{graphs}, if $C=(V,E,W)$, and $W:E \rightarrow \mathbb{R}^{+}_{*}$, the graph $\alpha C$ is defined as  $\alpha C=(V,E,W')$, where $W' = f \circ W$, with $f(x) = \alpha x$. 
In terms of resistive circuits, the individual resistances of the new circuit are multiplied by a constant $\alpha$ after the graph multiplication. It follows that to maintain consistency with the interpretation of circuits as graphs, we must assume that $\alpha>0$.
Diagrammatically, this operation will be represented as shown in figure~\ref{multiplication}.

We can now define the self-similar resistive series. The two parameters which characterize this configuration are: $R_{0}$, the resistance of the initial circuit $C_{0}$, and a positive real number $\alpha$. As we will see in the following, $\alpha$ must be non-null and smaller than $1$, and can be interpreted as an ``attenuation parameter''. For $\alpha$ close to $1$, we have a small attenuation. For $\alpha$ close to $0$, the attenuation is significant.

\begin{figure}[h]

\centering
\includegraphics[width=0.5\columnwidth]{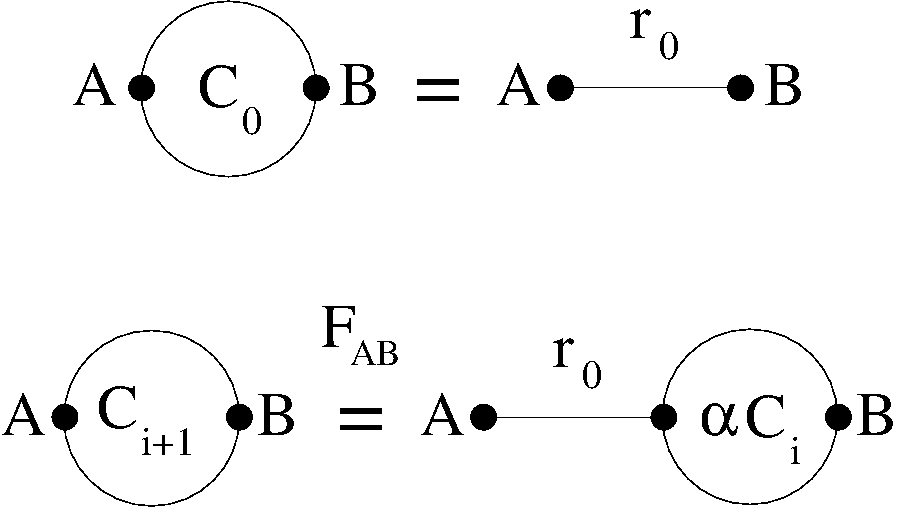}

\caption{First element $C_{0}$ and recurrence relation $C_{i+1} = F_{AB}(C_{i})$ for the self-similar resistive series. The same $C_{0}$ will be used in the next section in the definition of self-similar resistive trees.}
\label{C0_series_rec}
\end{figure}

\begin{figure}[h]
\centering

\includegraphics[width=0.65\columnwidth]{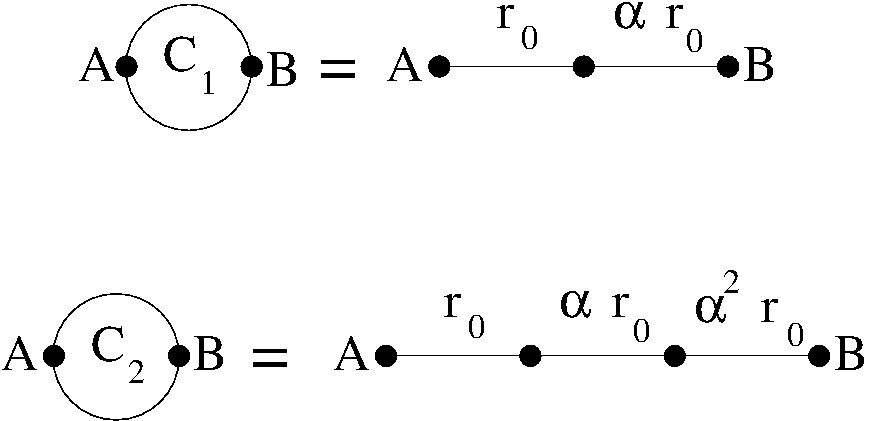}

\caption{Elements $C_{1}$ and $C_{2}$ of the sequence associated to the self-similar resistive series.}
\label{circuitoserie2}
\end{figure}

We start with the first element $C_{0}$ of the sequence, presented in figure~\ref{C0_series_rec} (left). The sequence $(C_{i})$ which defines the self-similar resistive series is given by the recurrence relation presented in figure~\ref{C0_series_rec} (right). For example, the second and third elements of the sequence $(C_{i})$ associated to the self-similar resistive series are shown in figure~\ref{circuitoserie2}.

A few words are in order concerning the self-similar resistive series. The first point is that, although this configuration cannot be readily associated to a geometric fractal, it is consistent with the general definition presented in section~\ref{general-definition}, and therefore is a self-similar structure. Also, it is one of the simplest self-similar resistive network, illustrating the main characteristics of the general definition with minimum technical difficulties. 

\begin{figure}[h]

\centering
\includegraphics[width=0.7\columnwidth]{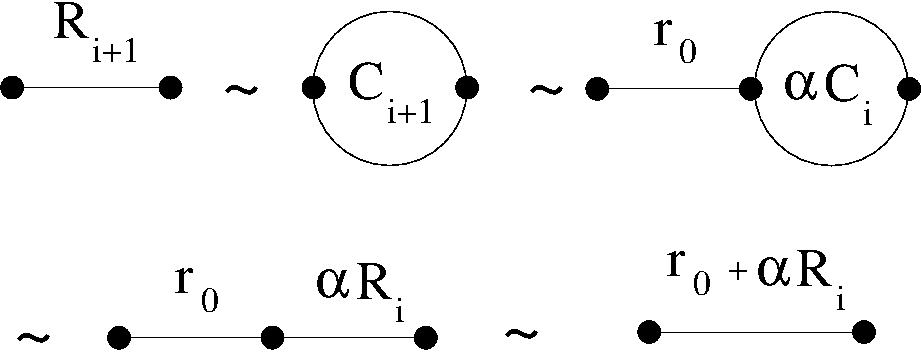} 

\caption{Diagrammatic representation for the development towards equation~(\ref{iter12}).}
\label{series_development}
\end{figure}

We now focus on the calculation of the equivalent resistance. The first element of the self-similar sequence is given in figure~\ref{C0_series_rec}, so
\begin{equation}
R_{0} = r_{0} \, .
\end{equation}
Applying $R$ to the recurrence relation presented in figure~\ref{C0_series_rec}, resistance  $R_{i+1}$ is obtained from $R_{i}$. Following  the development presented diagrammatically in figure~\ref{series_development}, and using the series transformation presented in figure~\ref{series-parallel}, we obtain
\begin{equation}
R_{i+1} = r_{0} + \alpha R_{i} \, .
\label{iter12}
\end{equation}
From equation~(\ref{iter12}), it follows that the recurrence function $T$ for the sequence $(R_{i})$ is 
\begin{equation}
T(x) = R_{0} + \alpha x \, .
\label{T_serie}
\end{equation}
It is straightforward to determine conditions for the sequence $( C_{i} )$ to represent a self-similar resistive circuit. From equation~\ref{T_serie}, 
\begin{eqnarray}
d(T(x),T(y)) & = &  \left| T(x) - T(y) \right| \nonumber \\
& = &  \left| R_{0} + \alpha x - (R_{0} + \alpha y) \right| \nonumber \\
& = & \alpha \left| x-y \right| \nonumber \\
& = & \alpha d(x,y) \, .
\end{eqnarray}
As discussed in section~\ref{general-definition}, the sequence $(R_{i})$ is convergent if $0 \le \alpha<1$. Therefore, the recurrence relation defined in figure~\ref{C0_series_rec} produces a self-similar circuit if the attenuation parameter is in the range $0 < \alpha < 1$.

The equivalent resistance of the self-similar resistive series can now be obtained. Considering the fixed point equation~(\ref{Req}) for $0 < \alpha<1$, we have%
\begin{equation}
R_{eq} = R_{0} + \alpha R_{eq} \, .
\end{equation}
The equivalent resistance is then given by
\begin{equation}
R_{eq}=\frac{R_{0}}{1-\alpha} \,\,\, \textrm{with} \,\,\, 0 < \alpha<1 \, .
\label{eqserier}
\end{equation}
We see that $R_{eq}$ becomes arbitrarily large with $\alpha$ close to $1$, the limit where the resistor array becomes an infinite series of resistors with the same resistance. On the other hand, if the attenuation is large (that is, if $\alpha$ is small), the equivalent resistance of the self-similar circuit is close to the resistance of a single resistor.

\section{Self-similar resistive trees}
\label{tree-circuits}

In this section we consider more complex structures, analogous to fractal trees. 
More specifically, we will define circuits that, when one of the external vertex (with its associated edges) is removed, the resulting graph is a tree. In the terminology of Graph Theory, a tree is a connected graph with no cycles \cite{harary1969graph}.
Hence, a two terminal resistive tree is a circuit represented by a graph that have no internal cycles but the ones removed with the elimination of an external vertex.
The physical motivation for the introduction of the extra vertex and edges in the self-similar resistive tree is to ensure that the electric potential of the ``tree branches'' are the same. In the process, we transform a tree configuration into a two-terminal circuit, which can be treated with the general formalism introduced in section~\ref{general-definition}.%
\footnote{A visual picture of the proposed configuration would be an actual tree trunk conducting electricity up to the thinner branches and then to the atmosphere. We would have effectively a two-terminal circuit, with one terminal being the base of the trunk (connected to the earth), and the other all of the branches (electrically attached to the atmosphere).}

For the present development, another operation with graphs is needed: the elementary contraction \cite{harary1969graph}. A contraction of a graph $G$ is obtained by identifying two vertices $U$ and $V$ that are connected by an edge. 
This contraction operation on graphs should not be confused with contraction in the context of metric spaces, discussed in section~\ref{general-definition}.

\begin{figure}[h]
\centering

\includegraphics[width=0.8\columnwidth]{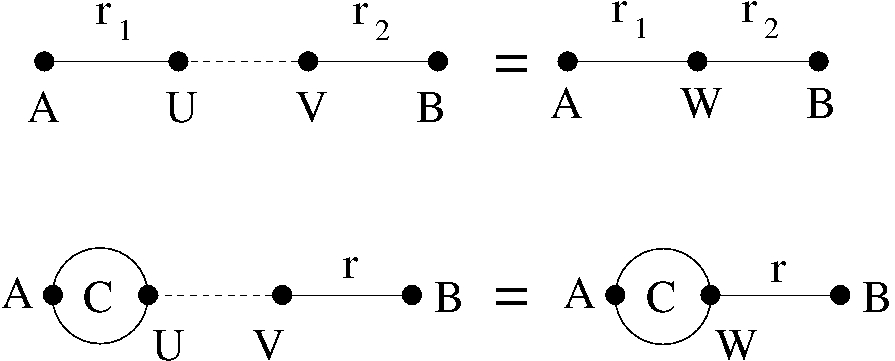}

\caption{Examples of the diagrammatic representation for the contraction of vertices $U$ and $V$ into vertex $W$. Dashed lines represent contractions.}
\label{contractions}
\end{figure}

Diagrammatically, we will denote a contraction by a dashed line. We illustrate this operation in figure~\ref{contractions}.
Physically, a contraction in a resistive circuit means that a resistor is substituted by a ideal connector. As we have seen in section~\ref{graphs}, the connector short-circuits two points of the circuit, which makes the two vertices in the associated graph to ``collapse'' into a single one.

\begin{figure}[h]
\centering
\includegraphics[width=0.8\columnwidth]{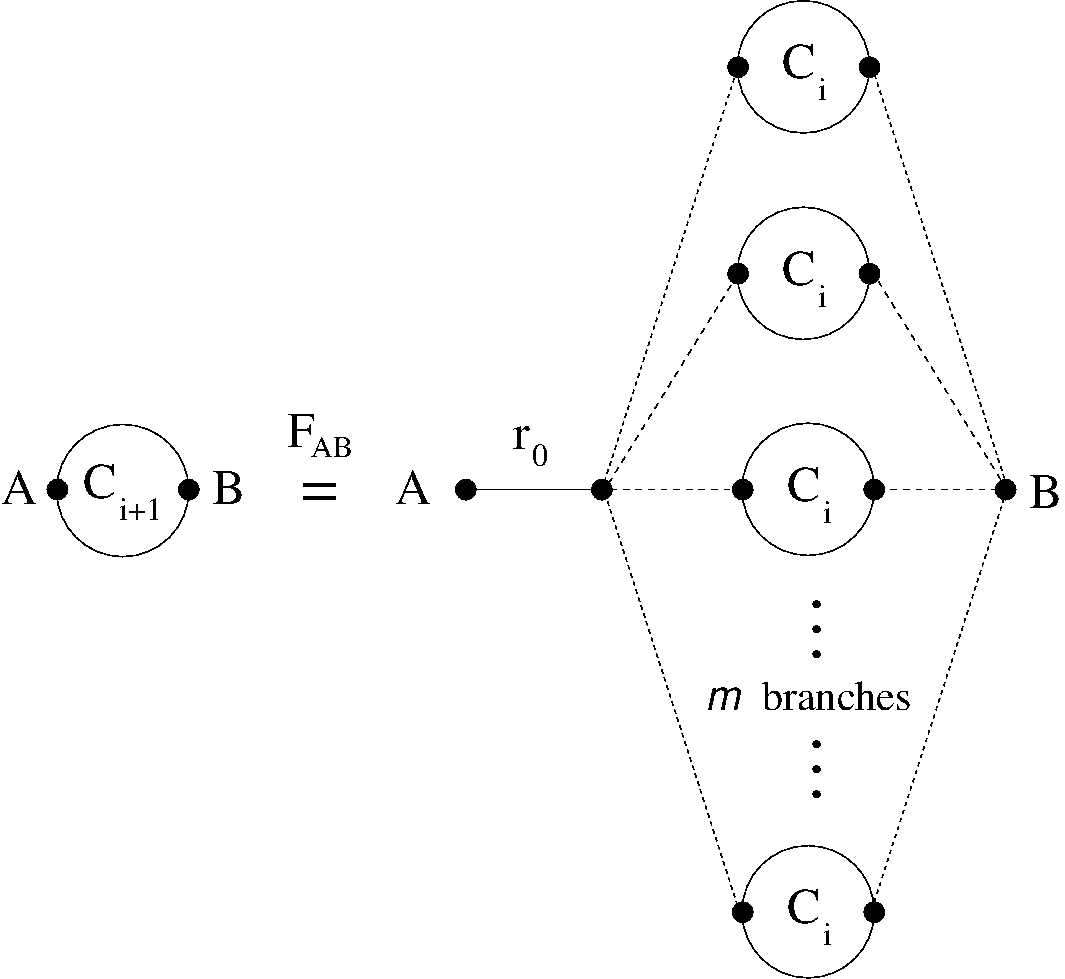}

\caption{Recurrence relation $C_{i+1} = F_{AB}(C_{i})$ for the self-similar resistive tree with $m$ branches. Dashed lines represent contractions.}
\label{recrelation_tree}
\end{figure}

\begin{figure}[h]

\centering
\includegraphics[width=0.8\columnwidth]{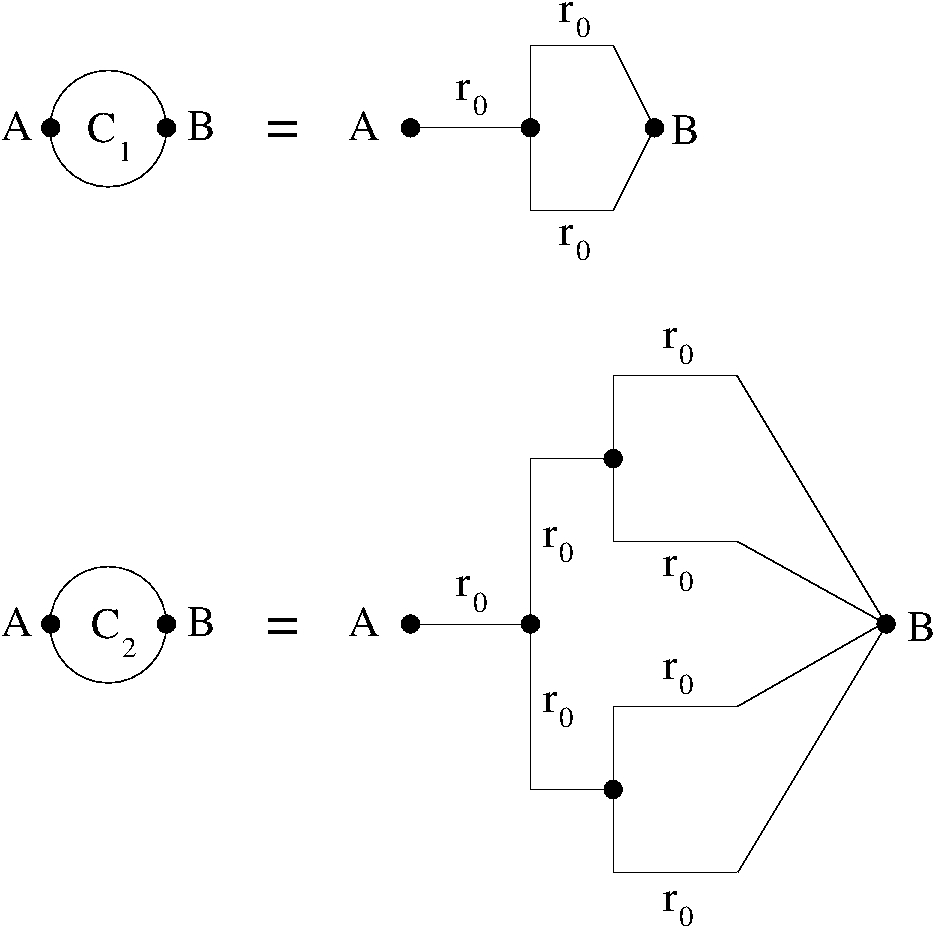}

\caption{Elements $C_{1}$ and $C_{2}$ of the sequence associated to the self-similar binary tree ($m=2$).}

\label{binary_tree}
\end{figure}

The parameters which characterize self-similar trees are the resistance $r_{0}$ of each resistor in the network, and the number of branches $m$ of the tree. We will require that $m>1$.
Starting from a single resistor, circuit $C_{0}$ in figure~\ref{C0_series_rec} (left), the construction of the self-similar tree is done following the recurrence rule indicated in figure~\ref{recrelation_tree}. As an illustration, we show the second and third elements of a resistive self-similar tree with two branches, a ``binary tree'', in figure~\ref{binary_tree}.

Since the first element of the self-similar sequence is given in figure~\ref{C0_series_rec} (left), we have
\begin{equation}
R_{0} = r_{0} \, .
\end{equation}
Applying series and parallel association (see figure~\ref{series-parallel}) with the recurrence relation in figure~\ref{recrelation_tree},  recurrence formula for the sequence $( R_{i} )$ is obtained,
\begin{equation}
R_{i+1} = r_{0} + \frac{R_{i}}{m} \, .
\label{iter3}
\end{equation}
The recurrence function associated to equation~(\ref{iter3}) is
\begin{equation}
T(x) = r_{0} + \frac{x}{m} \, ,
\end{equation}
and for $x,y \in \mathbb{R}^{+}_{*}$ we have
\begin{eqnarray}
d(T(x),T(y)) & = & | T(x) - T(y)| \nonumber \\
& = & \left| r_{0} + \frac{x}{m} - ( r_{0} + \frac{y}{m}) \right| \nonumber \\
& = & \frac{|x-y|}{m} \nonumber \\
& = & \frac{d(x,y)}{m} \, .
\label{development_tree}
\end{eqnarray}
Since the number of branches $m$ is greater than $1$, equation~(\ref{development_tree}) shows that in the present case $T$ is always a contraction. Hence, the sequence $( R_{i} )$ converges to some $R_{eq}$ and the sequence $( C_{i} )$ is associated to a well-defined self-similar resistive circuit.

The equivalent resistance $R_{eq}$ is the solution of the fixed point equation (\ref{Req}), 
\begin{equation}
R_{eq} = r_{0} + \frac{R_{eq}}{m} \, .
\end{equation}
Therefore, for the self-similar resistive tree with $m$ branches, we have
\begin{equation}
R_{eq}=\frac{m}{m-1} \, r_{0}  \,\,\, \textrm{with} \,\,\, m=2,3,\ldots \, . 
\end{equation}
We see that the largest equivalent resistance is obtained with two branches ($m=2$). The more $m$ grows, the closer $R_{eq}$ is to the initial resistance $r_{0}$.

\section{Sierpinski resistive circuits}
\label{Sierpinski-circuits}

\begin{figure}[h]

\centering
\includegraphics[width=0.6\columnwidth]{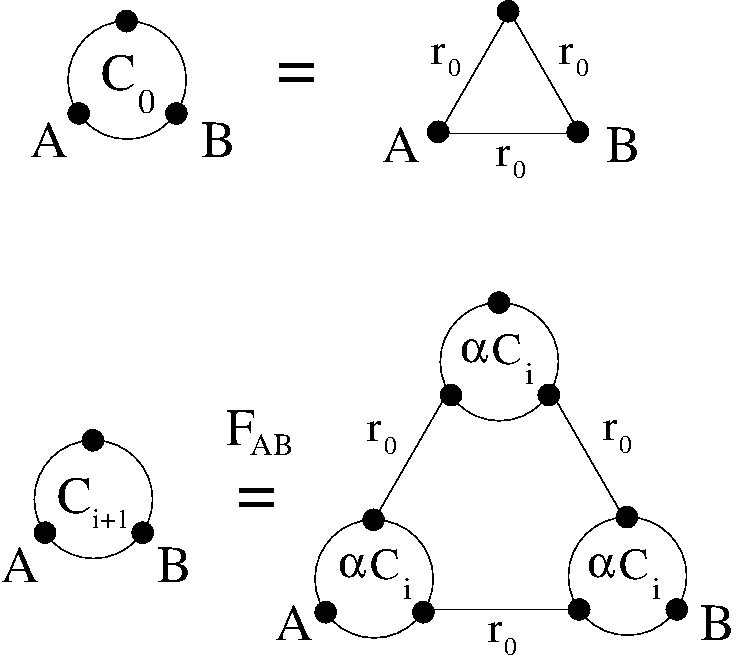}

\caption{First element $C_{0}$ and recurrence relation $C_{i+1} = F_{AB}(C_{i})$ for the Sierpinski self-similar resistive circuit.}
\label{C0_sierpinski}
\end{figure}

\begin{figure*}
\centering

\includegraphics[width=0.6\textwidth]{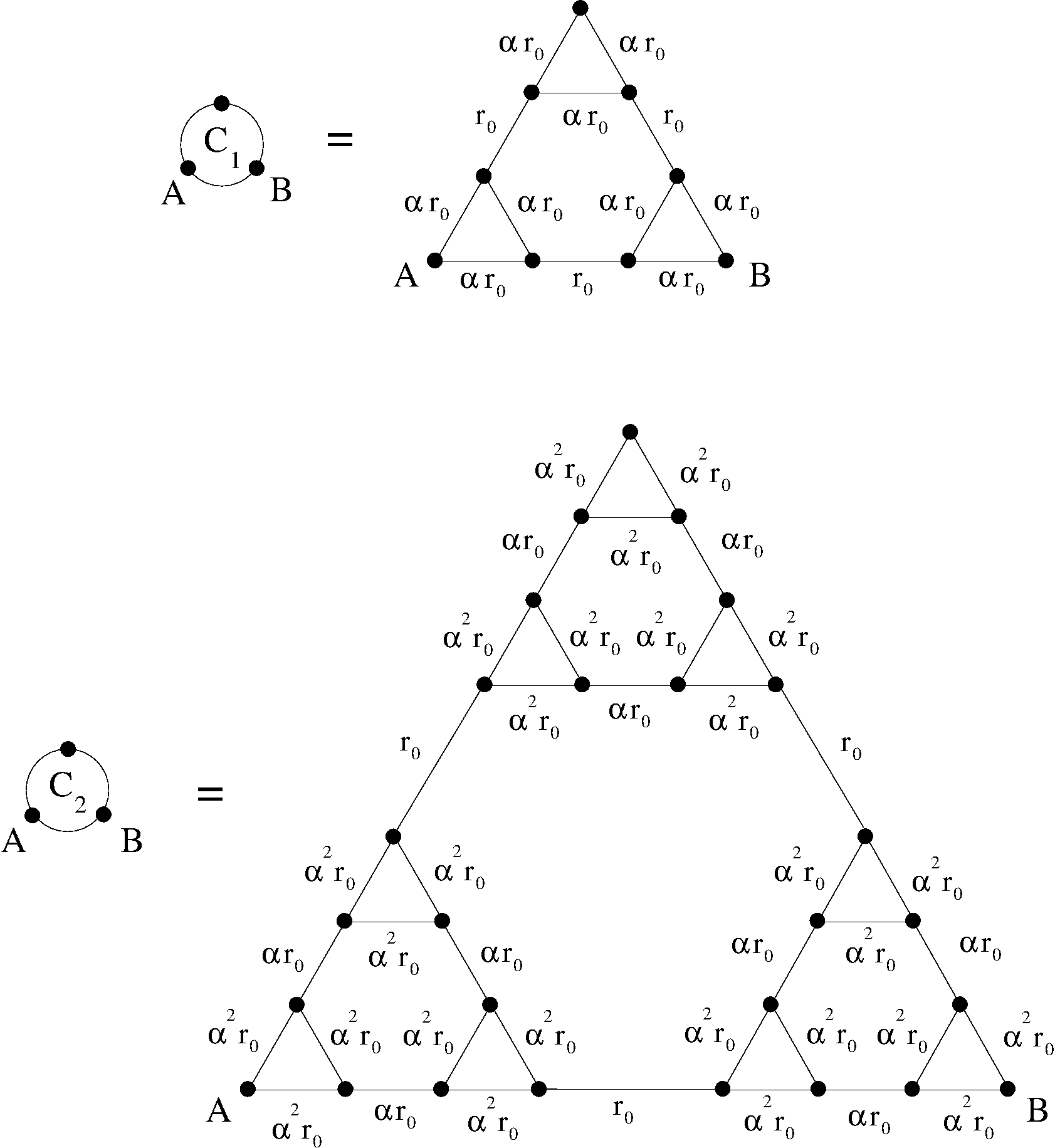}

\caption{Elements $C_{1}$ and $C_{2}$ of the sequence associated to the Sierpinski self-similar resistive circuit.}

\label{sierpinski}
\end{figure*}

A more complex class of circuits to be treated is inspired in the Sierpinski triangle \cite{kirillov2013tale,bedrosian,boyle,Alstrom198820}, a fractal also called Sierpinski gasket or Sierpinski sieve.
For the definition of the Sierpinski self-similar resistive configuration, we take the initial circuit $C_{0}$ and the recurrence function presented in figure~\ref{C0_sierpinski}. There are two parameters, $r_{0}$ and $\alpha$, for the Sierpinski circuit.
To illustrate the configuration, we show the second and third elements of the Sierpinski sequence in figure~\ref{sierpinski}.
 
Because of the relative complexity of the circuits involved, the determination of the recurrence function $T$ is not so straightforward as in the previous examples. Simple series and parallel association formulas are not enough. In the present case, the $Y-\Delta$ transform, presented in figure~\ref{Ydelta}, will be extensively used.

\begin{figure}[h]
\centering
\includegraphics[width=0.7\columnwidth]{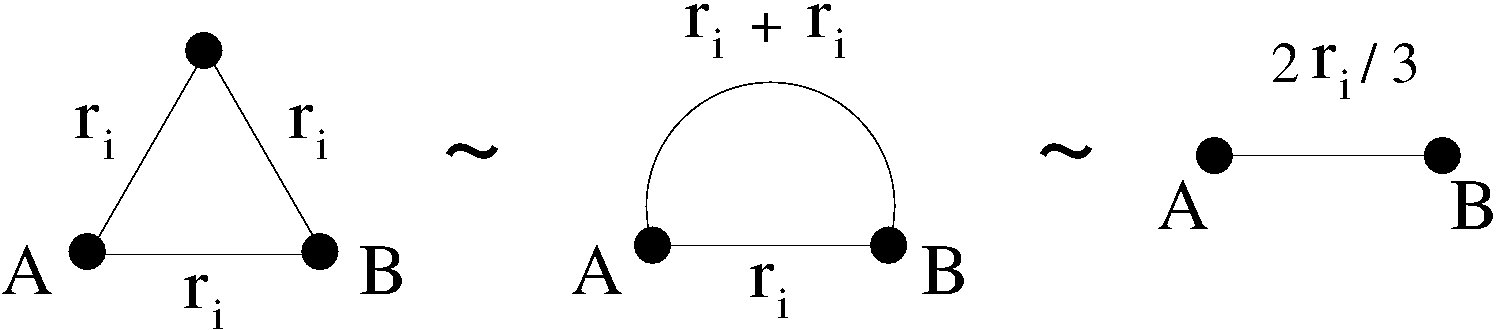}

\caption{Equivalent resistance of a $\Delta$-type circuit with individual resistances $r_{i}$ and external terminals $A$ and $B$.}
\label{sierpinski_development}
\end{figure}

The first result we comment is the equivalent resistance of a $\Delta$-type circuit, considering as external points two vertices of the triangle. The sequence of equivalent resistances used in the calculation is shown in figure~\ref{sierpinski_development}. The result is 
\begin{equation}
R_{i} = \frac{2}{3} \, r_{i} \, .
\label{R0_sierpinski}
\end{equation}
Equation~(\ref{R0_sierpinski}) can be immediately applied to $C_{0}$, the initial circuit of the Sierpinski sequence $(C_{i})$, and therefore $R_{0} = 2 r_{0} / 3$.

\begin{figure}[h]

\centering

\includegraphics[width=0.75\columnwidth]{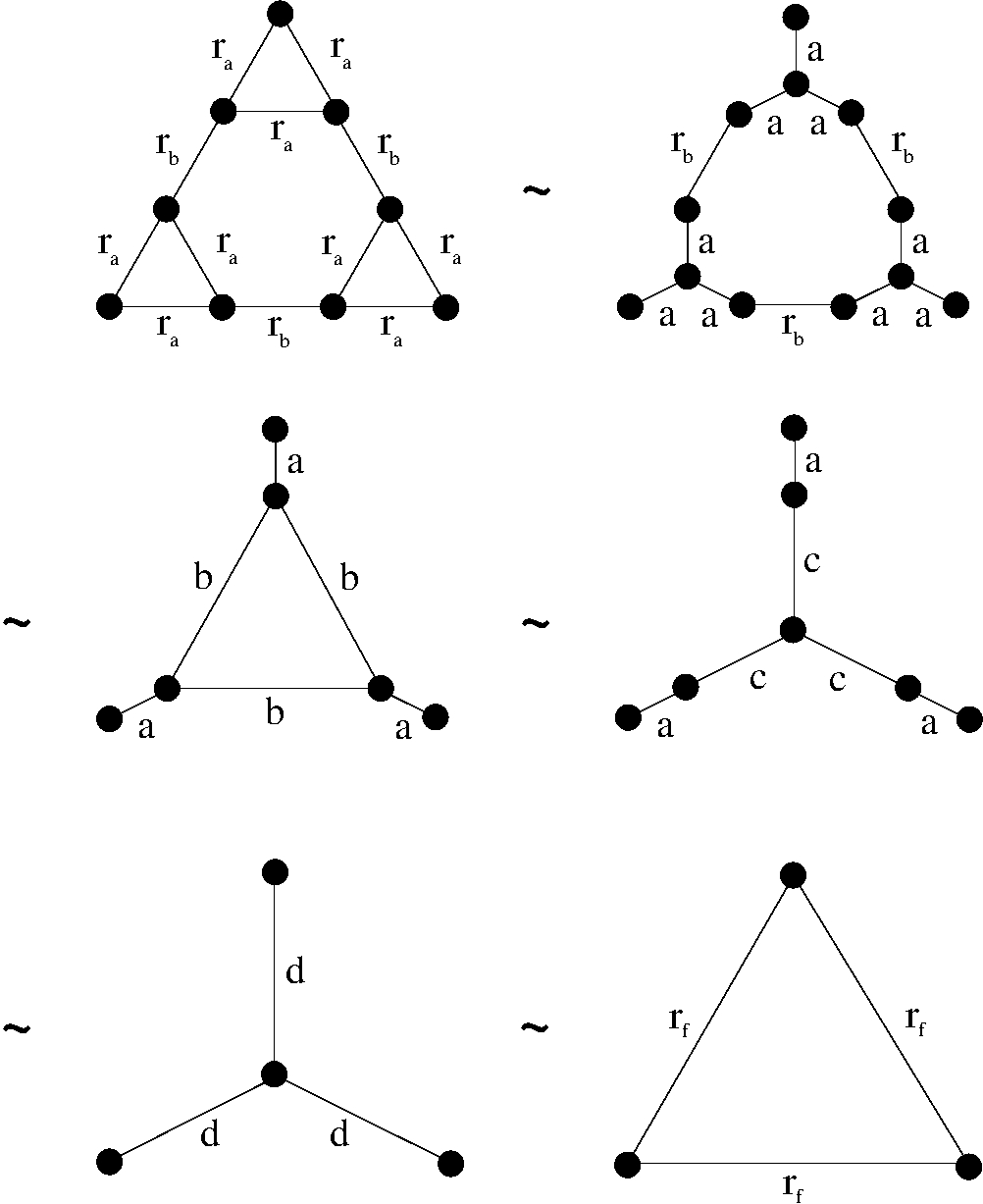}

\caption{Transformations that take one configuration presented in the Sierpinski sequence to a \mbox{$\Delta$-type} circuit with same equivalent resistance. In the indicated successive transformations: $a = r_{a}/3$, $b=2a + r_{b} = 2r_{a}/3 + r_{b}$, $c=b/3 = 2 r_{a}/9 + r_{b}/3$, $d=a+c=5r_{a}/9 + r_{b}/3$, $r_{f}=3d=5r_{a}/3 + r_{b}$.}

\label{sierpinski_moves}
\end{figure}

Another relevant result for the present development is displayed in figure~\ref{sierpinski_moves}. Using the $Y-\Delta$ transformation and the formula for association in series of resistors, we see that a composition of three $\Delta$-type circuits have the same equivalent resistance of a single $\Delta$-type circuit. The relation of the parameters $r_{f}$ and $r_{a}$, $r_{b}$, according to the calculations presented in figure~\ref{sierpinski_moves}, is
\begin{equation}
r_{f} = \frac{5}{3} \, r_{a} + r_{b} \, .
\label{relacao_sierpinski}
\end{equation}

Let us consider the form of recurrence relation for the Sierpinski sequence $(C_{i})$. We see that since we are starting from $C_{0}$, which is $\Delta$-shaped, the next circuit $C_{1}$ will be a combination of $\Delta$-type circuits in the arrange indicated by the first diagram in figure~\ref{sierpinski_moves}. 
The next circuit $C_{2}$ will not be a combination of $\Delta$-type circuits, but it will have an equivalent resistance of a combination of $\Delta$-type circuits. Notice now that the individual resistances of the $\Delta$-shaped subcircuits and the resistors linking them will be different in this case. Still, relation~(\ref{relacao_sierpinski}) can be applied. The same reasoning goes on for $\{ C_{3}, C_{4}, \ldots \}$. 

\begin{figure}[h]
\centering

\includegraphics[width=0.8\columnwidth]{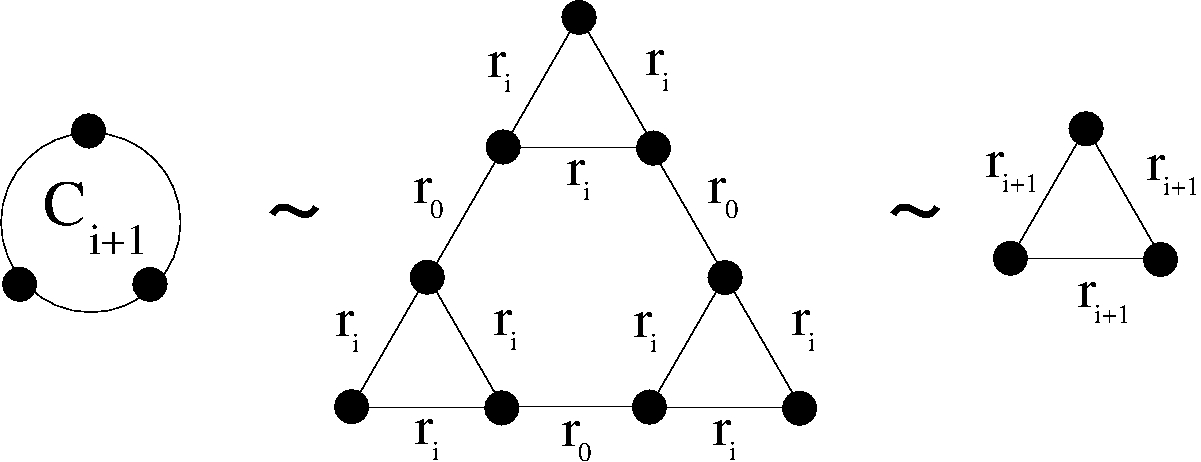}

\caption{Circuit $C_{i+1}$ has the same equivalent resistance of a combination of three $\Delta$-type circuits ($r_{i}$ and $r_{0}$ resistors), which has the same equivalent resistant of a single $\Delta$-type circuit ($r_{i+1}$ resistors).}
\label{sierpinski_equiv}
\end{figure}

The development presented in the previous paragraph is translated diagrammatically in figure~\ref{sierpinski_equiv}. Since the circuit $C_{i}$ has the same equivalent resistance of a $\Delta$-type circuit with resistors $r_{i}$, the circuit $C_{i+1}$ is equivalent to a $\Delta$-type circuit with resistors $r_{i+1}$, where
\begin{equation}
r_{i+1} = \frac{5}{3} \, \alpha r_{i} + r_{0} \, . 
\end{equation}
Also, using equation~(\ref{R0_sierpinski}), we obtain the recurrence relation for the sequence $(R_{i})$ of equivalent resistances,
\begin{equation}
R_{i+1} = \frac{5}{3} \, \alpha R_{i} + \frac{2}{3} \, r_{0} \, . 
\label{recurrence_resistances_sierpinski}
\end{equation}
From equation~(\ref{recurrence_resistances_sierpinski}) the recurrence function $T$ is \linebreak immediately read:
\begin{equation}
T(x) = \frac{5 \alpha}{3} \, x + \frac{2}{3} \, r_{0} \, .
\label{T_sierpinski}
\end{equation}

We can now determine if the sequence $(R_{i})$ is convergent. From equation~(\ref{T_sierpinski}) and considering $x,y \in \mathbb{R}^{+}_{*}$, we have
\begin{eqnarray}
d(T(x),T(y)) & = & | T(x) - T(y)| \nonumber \\
& = &
\left|
\frac{5 \alpha}{3} y + \frac{2}{3} \, r_{0} 
- \left(
\frac{5 \alpha}{3} x + \frac{2}{3} \, r_{0} 
\right)
\right| \nonumber \\
& = &  \frac{5 \alpha}{3} |y - x|\nonumber \\
& = &  \frac{5 \alpha}{3} d(y,x) \, .
\label{development_sierpinski}
\end{eqnarray}
Therefore, the sequence $(R_{i})$ is convergent if the attenuation parameter $\alpha$ is such that $0 < \alpha < 3/5$. In that case, the Sierpinski  recurrence relation in figure~\ref{C0_sierpinski} produces a well-defined self-similar circuit. 

Assuming that $0 < \alpha < 3/5$, it is possible to define and calculate an equivalent resistance for the self-similar Sierpinski circuit. Considering the fixed-point equation for $R_{eq}$,
\begin{equation}
R_{eq} = \frac{5 \alpha}{3} \, R_{eq} + \frac{2}{3} \, r_{0} \, , 
\end{equation}
we obtain the equivalent resistance, 
\begin{equation}
R_{eq} =  \frac{2r_{0}}{3 - 5 \alpha} \,\,\, \textrm{with} \,\,\, 0 < \alpha < \frac{3}{5}  \, .
\end{equation}
In the Sierpinski resistive circuit, $R_{eq}$ can be arbitrarily large, growing from $2 r_{0}/3$ to infinity as the parameter $\alpha$ varies from $0$ to $3/5$.

\section{Final comments}
\label{final-comments}

Resistor networks in self-similar patterns, analogous to geometric fractals, were considered in the present work. A main point to be stressed is that, since ideal resistive circuits are topological objects, geometric information should not be used in the construction of the configurations. 
To implement this idea, a precise definition for a self-similar resistive circuit was introduced. This definition captures essential characteristics presented in geometrical fractals, maintaining at the same time the topological nature of the circuits. 

In the definition of self-similar circuits discussed in the present work (but not universally adopted), the basic criterion is that the equivalent resistance of the self-similar network should be finite. 
This extra requirement is physically motivated. Since the equivalent resistance of an ideal resistive circuit is the most simple and important observable, it should be well-defined. 
Following this observation, we introduced a sufficient condition for the sequence to be a self-similar resistive circuit. That condition was the existence of a contraction for the sequence of associated resistances of each element in the sequence of circuits. If the criterion is satisfied, then the problem of calculating the equivalent resistance becomes a fixed point problem.

The approach presented in this work was illustrated in the construction of three classes of self-similar resistive circuits, discussed in detail. The first configuration, the self-similar series of resistors, is a simple and pedagogical case, where the formalism can be developed with minimum technical difficulty. 
Tree-shaped configurations and circuits based on the Sierpinski triangle
were considered next. Those more elaborate networks have some of the general characteristics presented by geometric fractals and, at the same time, they illustrate some of their important properties.

In this manuscript, our main goal was to introduce an interesting and non-trivial problem which could be used in Physics and Mathematics teaching. Nevertheless, applications in materials science and engineering are expected. To cite just a few possible scenarios, electrical properties of percolation clusters in random media and disordered systems can be studied considering fractal networks \cite{Clerc}. Sierpinski gasket can be used to model two dimensional superconductor materials \cite{Taitelbaum}. The electric response of inhomogeneous materials can be investigated with fractal-like models \cite{Clerc2}. Alternative antenna designs modeled by self-similar structures were considered \cite{Gopalakrishnan}. 

Finally, the particular examples explored in this work can be generalized. For instance, it is straightforward to introduce an attenuation factor in the definition of the tree circuit. In addition, the Sierpinski triangle circuit can be modified to a pattern similar to Sierpinski carpet \cite{Elvia,kirillov2013tale}. Self-similar circuits constructed from the general definition presented here can have interesting patterns and potentially important applications.

\begin{acknowledgments}
C. X. M. S. acknowledges the support of CAPES, Brazil. 
C. M. M. acknowledges the support of FAPESP (grant No.~2015/24380-2) and CNPq (grant No.~307709/2015-9), Brazil.
\end{acknowledgments}


\begin{thebibliography}{99}                             

\bibitem{Camp} Dane R. Camp, Mathematics Teacher, \textbf{84}, 265 (1991).

\bibitem{Portes} Leonardo L. Portes, Revista Brasileira de Ensino de F\'{\i}sica \textbf{21}, 501 (1999). 

\bibitem{Amaku} M. Amaku, M. Moralles, L. B. Horodysnki-Matsushigue, P. R. Pascholati, Revista Brasileira de Ensino de F\'{\i}sica \textbf{23}, 422 (2001).   

\bibitem{Elvia} \'{E}lvia Mureb Sallum, Revista do Professor de Matem\'{a}tica \textbf{57}, 1 (2005).

\bibitem{Assis} Thiago Albuquerque de Assis, Jos\'{e} Garcia Vivas Miranda, Fernando de Brito Mota, Roberto Fernandes Silva Andrade, Caio M\'{a}rio Castro de Castilho, Revista Brasileira de Ensino de F\'{\i}sica \textbf{30}, 2304 (2008).

\bibitem{Caicedo} H. E. Caicedo-Ortiz, H. O. Castañeda, E. Santiago-Cort\'{e}s, Revista Brasileira de Ensino de F\'{\i}sica \textbf{39}, e3308 (2017).

\bibitem{mandelbrot2004fractal} B. B. Mandelbrot, M. L. Lapidus, M. Van Frankenhuysen, \textit{Fractal Geometry and Applications: Analysis, number theory, and dynamical systems} (American Mathematical Society, 2004).

\bibitem{mandelbrot1983fractal} B. B. Mandelbrot, \textit{The Fractal Geometry of Nature} (Henry Holt and Company, 1983).

\bibitem{kirillov2013tale} A. A. Kirillov, \textit{A Tale of Two Fractals} (Springer, 2013).

\bibitem{zemanian} A. H. Zemanian, IEEE Transactions on Circuits and Systems \textbf{35}, 1346 (1988).

\bibitem{atkinson} D. Atkinson, F. J. van Steenwijk, American Journal of Physics \textbf{67}, 486 (1999).

\bibitem{cserti} J\'{o}zsef Cserti, Gyula D\'{a}vid, Attila Pir\'{o}th, American Journal of Physics \textbf{70}, 153 (2002). arXiv: cond-mat/0107362.

\bibitem{hansen:1986} Alex Hansen, Mark Nelkin, Physical Review B \textbf{33}, 649 (1986).

\bibitem{chen2016} Joe P. Chen, Luke G. Rogers, Loren Anderson, Ulysses Andrews, Antoni Brzoska, Aubrey Coffey, Hannah Davis, Lee Fisher, Madeline Hansalik, Stephew Loew, Alexander Teplyaev, Power dissipation in fractal AC circuits, arXiv:1605.03890 (2016).

\bibitem{ruiz2017} Patricia Alonso Ruiz, Power dissipation in fractal Feynman-Sierpinski AC circuits, arXiv:1701.08039 (2017).

\bibitem{bedrosian} Samuel D. Bedrosian, Xiaoguang Sun, Journal of the Franklin Institute \textbf{326}, 503 (1989).

\bibitem{boyle} Brighid Boyle, Kristin Cekala, David Ferrone, Neil Rifkin, Alexander Teplyaev, Pacific Journal of Mathematics \textbf{233}, 15 (2007).

\bibitem{Alstrom198820} Preben Alstrom, Dimitris Stassinopoulos, H. Eugene Stanley, Physica A \textbf{153}, 20 (1988).

\bibitem{Essoh:1988bn} C. D. Essoh, J. Bellissard, Journal of Physics A  \textbf{22}, 4537 (1989). 

\bibitem{nilsson2008electric} J. W. Nilsson, S. A. Riedel, \textit{Electric Circuits} (Pearson/Prentice Hall, 2008). 

\bibitem{harary1969graph} F. Harary, \textit{Graph Theory} (Addison-Wesley Publishing Company, 1969).  

\bibitem{haaser1991real} N. B. Haaser, J. A. Sullivan, \textit{Real Analysis} (Dover Publications, 1991). 

\bibitem{Clerc} J. P. Clerc, G. Giraud, J. M. Laugier, J. M. Luck, Journal of Physics A  \textbf{18}, 2565 (1984). 

\bibitem{Taitelbaum} H. Taitelbaum, S. Havlin, Journal of Physics A  \textbf{21}, 2265 (1988). 

\bibitem{Clerc2} J. P. Clerc, G. Giraud, J. M. Laugier, J. M. Luck, Advances in Physics \textbf{39}, 191 (1990).

\bibitem{Gopalakrishnan} Gopalakrishnan Srivatsun, Sundaresan Subha Rani, Gangadaran Saisundara Krishnan, Wireless Engineering and Technology \textbf{2}, 107 (2011).


\end{thebibliography}
\end{document}